\documentstyle[12pt,epsfig]{article}
\def\be{\begin{equation}}
\def\ee{\end{equation}}
\def\bea{\begin{eqnarray}}
\def\eea{\end{eqnarray}}

\def\e{{\rm e}}
\def\eps{\varepsilon}
\def\pt{{\rm PT}}

\begin{document}
\vspace*{4cm}

\begin{flushright}
YITP-SB-00-66\\
BNL-HET-00/40
\end{flushright}

\bigskip

\begin{center}

%\title
{\Large \bf QCD THEORY AT HIGH ENERGY \footnote{Based on a
talk presented at the {\it IVth Rencontres du Vietnam,
International Conference on Physics at Extreme Energies},
Hanoi, Vietnam, July 19-25, 2000.}}

\vbox{\vskip 0.25 true in}
%\author
{ George Sterman }

\vbox{\vskip 0.2 true in}
%\address
%\begin{center}
{\it C.N.\ Yang Institute for Theoretical Physics,
State University of New York at Stony Brook\\
Stony Brook, NY 11794-3840 USA\\
\ \\
Physics Department, Brookhaven National Laboratory\\
Upton, NY 11973, USA}
\end{center}

\vbox{\vskip 0.4 true in}

\begin{abstract}
{The energy range and quality of strong-interaction
data from recent years demand the study
of higher orders in perturbative QCD, 
and of nonperturbative effects. 
I discuss a selection of recent progress in the theory of
QCD at high energy, including examples from perturbative resummation, 
nonperturbative power corrections 
and the Regge limit.  In each case, techniques of
factorization play a central role.}
\end{abstract}

\section{Introduction}

Current studies in QCD are motivated partly
by its importance in the production
and detection of new physics, but also,
and in very large part, by the challenges of 
quantum chromodynamics itself.  
QCD may be thought of
as an exemplary quantum field theory, exhibiting asymptotic
freedom, confinement, chiral symmetry breaking and so on,
the workings of which are all available, and unavoidable,
within present energy ranges.  It is a vast subject, encompassing
perturbative, heavy-quark, nonrelativistic and lattice
QCD, all the way to nuclear physics.  In a very real sense,
what are sometimes called ``tests of QCD'' are 
tests of quantum field theory itself.

In this talk, I will discuss some recent
efforts to narrow the gap between the
high-energy, partonic and low-energy, hadronic descriptions
of QCD, starting from the high-energy side,
 through the study of higher-order corrections
in perturbation theory and of nonperturbative
power corrections.
 I will highlight the use of QCD factorization as an
organizing principle in these investigations.

The past decade has been a proverbial
golden age of hadronic data at high
energy, in terms both of coverage and of quality.  
There is no room here to do justice to the data revolution
of the 1990's, the work of a generation of accelerators
that has reached maturity: LEP, HERA and the Tevatron.
Each has produced spectacular successes for our current
picture of QCD, but each has provided  its share of puzzles.
Two examples must suffice.  Our control of
strong interaction corrections in inclusive cross sections
such as  deep-inelastic scattering (DIS) may be gauged
from Fig.~\ref{distotal}, showing data from HERA. \cite{ep}  
Here, the total ep
cross sections at momentum transfers from tens into  hundreds
of GeV track the standard model predictions, which  
include extensive input from perturbative QCD
evolution, described below.

\begin{figure}
\begin{center}
\hspace*{-7mm}
%\vspace*{3mm}
\epsfig{file=sfig8remnccc.epsi,width=6cm}%angle=90}
\end{center}
\caption{DIS cross section for
neutral and charged currents at high momentum transfers.
{\protect \cite{ep} }}
\label{distotal}
\end{figure}

\begin{figure}
\begin{center}
\hspace*{-7mm}
%\vspace*{3mm}
\epsfig{file=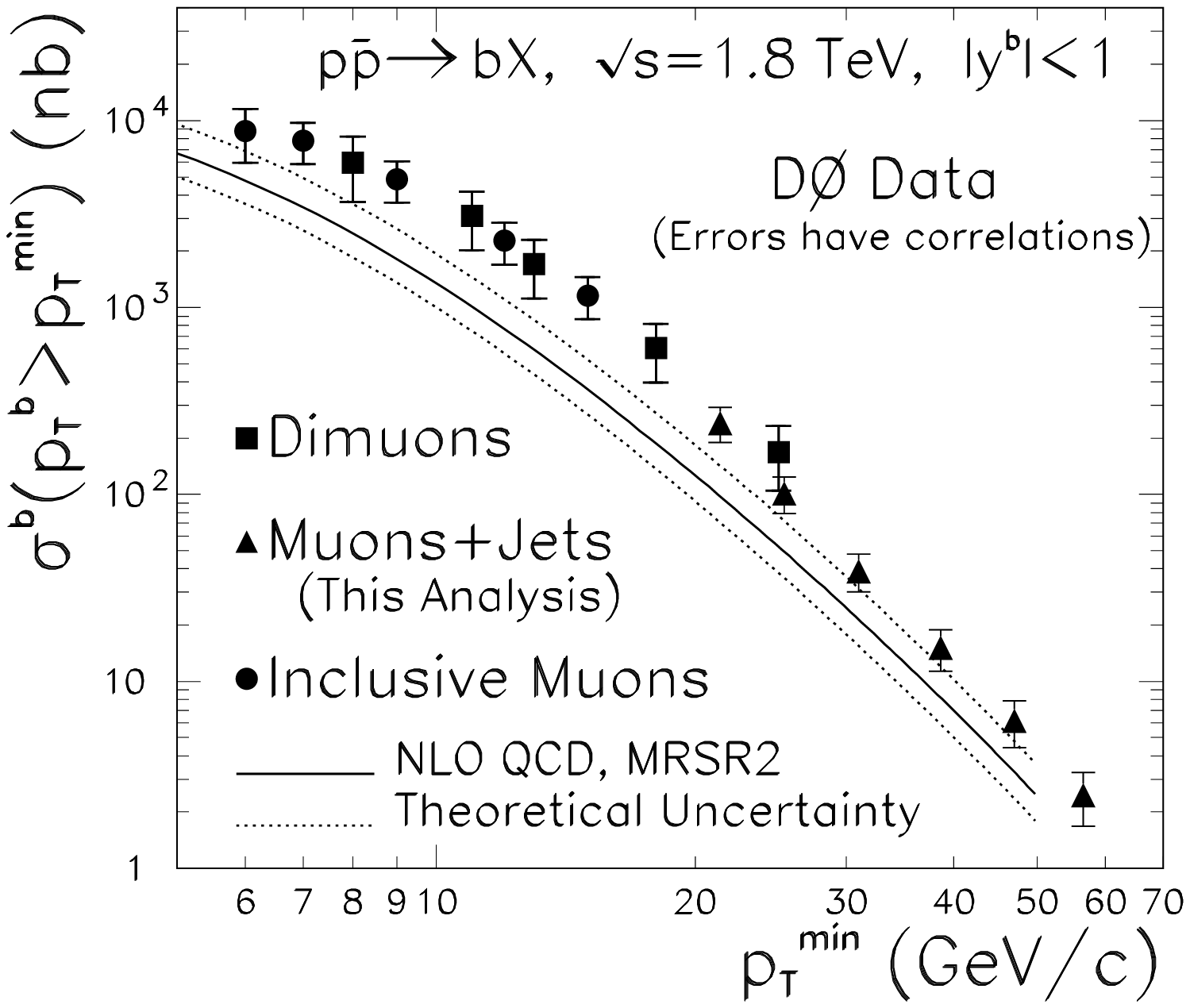,width=6cm}%angle=90}
\end{center}
\caption{Transverse-momentum distribution
for b-quark production at the Tevatron.{\protect \cite{bjetd0}}}
\label{bjetd0}
\end{figure}

On the other hand, Fig.\ \ref{bjetd0} shows
the transverse momentum distribution for b-quark production
at the Tevatron. \cite{bjetd0}   This data is 
typical of cases where our present theory is
partly adequate, partly not.  Overall, the theoretical $p_T$
spectrum has the correct shape, but the normalization
of the theory is too low at low $p_T$, even as it gradually
approaches the data at larger $p_T$.  Much of
the effort described below is aimed at using
such apparent discrepancies as  guides to the perturbative
and nonperturbative structure of the theory.
To see how, let us briefly review 
the elements of our present theoretical framework for
high-energy QCD, based on factorization.

\section{The Unity of QCD Factorizations}\label{subsec:fig}

The application of perturbation theory to high energy QCD begins with
asymptotic freedom and infrared safety, illustrated
by the perturbative expansions of the total cross
sections for $\rm e^+e^-$ annihilation to hadrons,
and to  final-state jets:
\begin{eqnarray}
Q^2\, \sigma(Q^2,\mu^2,\alpha_s(\mu))
=
\sum_n c_n(Q^2/\mu^2)\; \alpha_s^n(\mu) + {\cal O}(1/Q^p)
= \sum_nc_n(1)\; \alpha_s^n(\mu)+ {\cal O}(1/Q^p)\, ,
\label{afis}
\end{eqnarray}
where the $c_n$ are dimensionless
coefficients.  For an infrared safe cross
section,  the $c_n$ are free of dependence
on fixed mass scales (such as light quark masses),
which are absorbed into corrections that are
suppressed by some power, $p$, of the c.m.s.\ energy, $Q$.
In  most cases, the
$c_n$ are fully known only for a few, low orders.
Because the cross
section is a physical observable, it is
independent of the renormalization scale, $\mu$,
which can therefore be chosen to equal $Q$.  
For an asymptotically free theory,
the larger is $Q$, the better  any
 finite-order approximation becomes.

Relatively few cross sections are quite this
simple, however, but whenever a reaction involves
a scattering at large momentum transfer,
or the creation or decay of a heavy state,
we may isolate its short-distance components,
which can be treated 
perturbatively, from its long-distance, nonperturbative
components.  This is a procedure known as
{\it factorization}, which generalizes the
operator product expansion.  

For a factorized cross section,
Eq.\ (\ref{afis}) is replaced by an expression
of the general form,
\begin{eqnarray}
Q^2\sigma(Q,x)
=
\omega(Q/\mu,x/\xi,\alpha_s(\mu))\;
\otimes\;
f(\xi,\mu)
+{\cal O}(1/Q^p)
\, ,
\label{fact}
\end{eqnarray}
with a `` hard-scattering'', or
coefficient, function $\omega$, which is short-distance and perturbative,
in convolution with 
a ``soft'' function $f$, which is long-distance and nonperturbative.
In DIS, with $q$ the momentum transfer,
$x=2p\cdot q/Q^2$, but more generally
it represents any dimensionless ratios of large
momentum scales.  
The dimensional variable $\mu$ is the factorization scale, 
separating long and short distances.
As in (\ref{afis}), the physical cross
section is independent of $\mu$.  
In DIS, the soft function $f$ 
is a parton distribution function (PDF).
For hadron-hadron scattering, we have
two PDF's in convolution form.  In these cases,
the convolution in (\ref{fact})
is in terms of fractional momenta, $\xi$,
of the (one or more) partons that initiate
the hard-scattering process. 
Factorization is more general
than this, however, and we shall encounter other examples below.
Nearly always, the soft function can be 
interpreted as the  matrix element of some
(usually nonlocal) operator in QCD.

The basis of factorization is
always the quantum-mechanical incoherence
of dynamics at very short distances from
that at long distances, and, in Minkowski
space, the mutual incoherence of the 
dynamics of particles whose relative
velocity approaches the speed of light.

Whenever there is factorization, there is {\it evolution},
a consequence of the independence of the physical
cross section from the factorization scale,
\be
\mu{d\over d\mu} \ln \sigma(Q,x,m)
=
\mu{d\over d\mu} \ln\left\{\omega(Q/\mu,x/\xi,\alpha_s(\mu))\otimes 
f(\xi,\mu) \right\}=0\, .
\ee
Because $f$ and $\omega$ have in common
only the parton momentum fraction and $\alpha_s$, separation-of-variable 
arguments imply complementary equations for $f$ and 
 $\omega$:
\be
\mu{d  f(\xi,\mu)\over d\mu}=  P(z,\alpha_s) \otimes f(\xi/z,\mu) 
\quad {\rm and} \quad 
\omega(Q/\mu,\eta z,\alpha_s) \otimes P(z,\alpha_s) = - 
\mu{d  \omega(Q/\mu,\eta,\alpha_s) \over d\mu}\, ,
\ee
in terms of convolutions with splitting functions $P(z,\alpha_s)$.
The first of these ``DGLAP evolution" relations enables us to 
take PDFs determined at some reference scale, $\mu_o$,
and extrapolate to higher, or
lower, scales, wherever the running coupling is
not too large.  

Measurements of the strong
coupling based on these methods give $\alpha_s(M_Z) \sim 0.12$,
which suggests that at around 100 GeV,
  ${\cal O}(\alpha_s^2)$ is about one percent.
This is the nominal level of accuracy to which 
perturbative QCD may aspire at ``next-to-next-to-leading" (NNLO)
order.  Referring to Eq.\ (\ref{afis}),
the $c_n$'s are known to NLO ($n=1$) for
cross sections with up to four jets in $\e^+\e^-$. \cite{ng98}
To date, complete
NNLO calculations are available only for one-scale problems: 
the total $\e^+\e^-$ cross section, DIS and Drell-Yan.
Two loops are the current frontier for finite-order perturbative QCD,
and the past year has seen significant progress toward
the exact computation of two-loop scattering amplitudes
and coefficient functions $\omega$.  \cite{BDK}
At the same time, to use two-loop coefficient functions,
it will be necessary to have the splitting functions 
$P$ at {\it three} loops; and here also important
progress has been reported  within just the past
few months. \cite{LRV,KPS}  

Beyond exact calculations, DGLAP evolution is the
first among a set of methods that enable us to probe
properties of QCD perturbation theory to all orders.
Each of these methods is based upon the separation
of dynamics at different length scales.

\section{Resummation for Inclusive and Exclusive Cross Sections}

Hard-scattering, such as jet, heavy quark
and high-$p_T$ photon production, depends on a complex combination
of the long-distance dynamics of the external
hadrons, the short-distance perturbative subprocess
and the properties of final states
included in the cross section.
Using concepts of factorization, we are learning to 
treat a widening variety of such cross sections, and 
increasingly to compute 
classes of higher-order corrections.  We describe below three
applications of current interest.

\subsection{Partonic Threshold}

Our first example is so-called threshold
resummation, which applies to inclusive
hard-scattering hadronic cross
sections $AB\rightarrow F+X$, with
$F$=$\gamma^*$, W, Z, jets, heavy quark,
etc.\ of invariant mass $Q$.  We are interested in higher-order
corrections to the perturbative hard-scattering
function $\omega_{ab\rightarrow F}$ in
factorized cross sections, Eq.\ (\ref{fact}),
at {\it partonic threshold},  $z=Q^2/\hat{s}\to 1$, where the
partons $a$ and $b$ have just enough c.m.s.\ energy $\sqrt{\hat s}$
to produce the observed final state.  In the
threshold region, there is an incomplete cancellation
of emission and virtual radiative corrections,
which leads to  singular corrections of the 
form
\bea
\omega_{ab\to F}^{(r)}(z) &\sim& \left(C_a{\alpha_s\over 
\pi}\right)^r {1\over r!}
\; \left[{\ln^{2r-1}(1-z)\over 1-z}\right]_+\, 
\eea
at $r$th order,
with $C_q=C_F$, $C_g=C_A$.  In effect, for $z\rightarrow 1$,
there are two hard scales, $Q$ and $(1-z)Q$.
Because these singular distributions
arise from soft-gluon radiation, sensitive to the
scale $(1-z)Q$, but not to the scale $Q$, they
may be ``refactorized" from the hard-scattering
into functions that depend only on the flow of
color at a truly short-distance hard-scattering, which is
sensitive only to $Q$.
As for Eq.\ (\ref{fact}), the new factorization
implies a new evolution equation, from
which we derive a  resummation in moment space,
\bea
\tilde \omega_{ab\rightarrow F}(Q/\mu,N)
&\equiv& \int_0^1 dz z^{N-1} \omega_{ab\to F}(Q/\mu,z)
\nonumber\\
&=&
\exp\left[ -\int_0^1 dz{z^{N-1}-1\over 1-z} 
\int_{(1-z)^2Q^2}^{\mu^2}{d m^2\over m^2} A_{ab}(\alpha_s(m))
\, \right]\, ,
\label{omegamoment}
\eea
where $A_{ab}=(C_a+C_b)(\alpha_s/\pi)+\dots$ 
is an expansion in $\alpha_s$.

Threshold resummation is currently being explored
for most of the basic inclusive cross sections. \cite{trrev,flresum}
An important
consequence is a reduction of factorization-scale dependence
in resummed cross sections.
In NLO calculations, dependence on $\mu$ begins at 
the next order, ${\cal O}(\alpha_s^2$),
but often this residual sensitivity is uncomfortably
large.  In the resummed function Eq.\ (\ref{omegamoment}), however,
the $\mu$-dependence is determined by
\be
{d\ln \tilde \omega_{ab}(N,\mu) \over d\ln\mu} \sim {A_{ab}(\alpha_s)\; \ln N} 
\sim -  {d\ln [\tilde f_{a/A}(N,\mu)\; \tilde f_{b/B}(N,\mu)]\over d\ln\mu}\, ,
\label{rgomega}
\ee
where to the right the
$\tilde f$'s are moments of the parton distributions. 
Factorization requires that the function $A_{ab}$ that appears 
in Eq.\ (\ref{omegamoment})
is exactly the same as  the sum of the $\ln N$ terms in the moments
of the splitting functions $P_{aa}$ and $P_{bb}$.  This
leads to a very significant decrease in sensitivity to
the factorization scale compared to previous, fixed order
calculations. \cite{scale,thrphen}
This is
only the beginning of applications of threshold resummation, however, and we
anticipate important applications to the determination of parton
distributions and to the improvement of predictions for new particle
production.

\subsection{Power Corrections: Universality and Beyond}

Jet cross sections in $\rm e^+e^-$
annihilation are defined by adjustable parameters,
whose variation mediates between fully inclusive and
nearly exclusive cross sections.  As such, they are 
ideal for testing and improving our understanding of
QCD at intermediate distances.

The most-studied examples involve
light-mass dijet pairs.  Dijet events, 
which dominate the annihilation cross
section at high energy, can be described
in terms of  event shapes.  Perhaps the
best known of these is the 
thrust, $T$.  The thrust of an $\rm e^+e^-$ event is determined
approximately by finding
an axis that maximizes the quantity
$T =1- (m_1^2+m_2^2)/Q^2$, where
$m_1$ and $m_2$ are the invariant
masses of the sums of all particle momenta within the two
hemispheres defined by this axis.
As $T\rightarrow 1$, the final state is
characterized by two well-collimated jets.
A number of other familiar event shapes
may be derived from jet masses in a similar way.

For light-mass dijets,  the relative velocities
of the two jets again insures that their dynamical
developments into the final state are mutually independent,
This independence results in factorization at the level
of cross sections, and leads, in much the same manner
as above, to evolution equations, and to
$T$-moments that are quite
similar to the $z$-moments of threshold resummation
in Eq.\ (\ref{omegamoment}),
\bea
\int dT\, T^{N-1}\, {d\sigma_{\rm PT}(T)\over dT}
&\sim&
  \sigma_{\rm tot}\; 
\exp\;
\left[- \int_0^1 dy{y^N-1\over 1- y} \int_{(1-y)^2Q^2}^{(1-y) Q^2}
{d k_T^2\over k_T^2}\; A_{q\bar q}(\alpha_s(k_T))\right]\, .
\label{Tresum}
\eea
Such resummed cross sections improve
the perturbative description of
differential cross sections, $d\sigma_{\rm PT}/de$, 
for a class of event shapes $e$, including the thrust.  \cite{cttw}
Nevertheless, for small $1-T$, that is, close
to the limit of massless jets, fits based on
Eq.\ (\ref{Tresum}) fail to
describe the data.  This is not surprising, because
the lighter the jets, the more dependent
on long times are their cross sections, and correspondingly
the more sensitive they are to nonperturbative effects.
The mass of the jet is a ``dial'' for tuning
the importance of nonperturbative dynamics.  

Although perturbation theory cannot predict true 
long-time behavior, it can give hints as to its
nature.  In Eq.\ (\ref{Tresum}) these hints come
from the running coupling.  When
the variable $y$
gets close enough to one, the running coupling
in the integrals in (\ref{Tresum}) diverges,
signalling a breakdown of perturbation theory.
This divergence is
associated with a region of fixed size in $k_T$,
independent of $N$.
We should think of this as an
ambiguity in perturbation theory, which
is resolved in the full theory by nonperturbative
information. \cite{irrrev,conf}  Since the perturbative range
of integration in Eq.\ (\ref{Tresum}) should
remain meaningful, the minimal modification
necessary to (\ref{Tresum}) is to 
replace the lower limit of the $k_T$-integral
with a nonperturbative parameter: $\alpha_0$, with
\begin{equation}
\alpha_0 \equiv {1\over\mu_0}\int_0^{\mu_0} dk\, \alpha_s(k)\, ,
\label{npparam}
\end{equation}
where $\mu_0$ is a conveniently chosen cutoff.
The quantity $\alpha_0$ has the interpretation
of the integral of the running coupling over the nonperturbative region.

Letting $e\equiv 1-T$, we can invert the transform
(\ref{Tresum}).   
The substitution (\ref{npparam}) then produces a simple
shift in the perturbative spectrum, \cite{KSmorDW}
\bea
{d\sigma(e) \over de} =
{d\sigma_\pt(e-\lambda_e/Q) \over de} +
{\cal O}\left({1\over e^2Q^2}\right)\, ,
\label{shifteq}
\eea
where we note that the relative size of the effect is
$\lambda_e/eQ$, and that corrections begin at $1/(eQ)^2$.
Similar considerations apply to any event shape $e$
that vanishes in the limit of light-like dijets.

This approach has been formalized and applied
in Refs.\ \cite{disp}.  
The integral of the running coupling, Eq.\ (\ref{npparam}),
is thought of as a fundamental, universal parameter.
Applications, which include the
partial incorporation of higher-order perturbative terms,
provide an improved picture of differential
event shapes, and a unified 
description of low moments of $e=1-T\dots$, such as  $\int dT (1-T) d\sigma/dT$.
In addition, the
approximation relates first to second moments:
\begin{equation}
\langle e^2 \rangle = \langle e^2 \rangle_{\rm PT} + 2{\lambda_e\over Q}\; 
\langle e\rangle_{\rm PT}+{\lambda_e^2\over Q^2}\, .
\end{equation}
This description has been reasonably successful
in tying together the first moments of different event
shapes. \cite{eventfit}  
Second moments, however, show  large $1/Q^2$
corrections, compared to the predictions of 
strong-coupling universality, Eq.\  (\ref{npparam}).  \cite{l3}

A more general formalism involves the introduction of
nonperturbative ``shape'' functions \cite{KSKT}
that generalize Eq.\  (\ref{shifteq}) to 
a distribution of shifts due to soft radiation,
\bea
{d\sigma \over de} =
\int_0^{eQ} d\epsilon\; f_e(\epsilon)\; {d\sigma_\pt(e-\epsilon/Q) \over
de} +{\cal O}\left({1\over eQ^2}\right)\, .
\label{shape}
\eea
The most important features of this expression are
the level of the corrections, down by a full power
of $Q$ compared to Eq.\ (\ref{shifteq}), and 
the independence of the shape function 
 $f_e$  from $Q$.  The latter implies that
a fit to $f_e$ at
one value of $Q$, most conveniently  $Q=m_Z$,
is sufficient to predict the differential
cross section for all $Q$. \cite{KSKT}

Equation (\ref{shape}) may
be derived from the factorization properties of
soft radiation in perturbation theory, and 
the shape function itself has the interpretation
of a matrix element in QCD.  To be specific, 
in the case of thrust the matrix element 
is
\bea
f_{1-T}(\eps,\mu_{\rm IR})=
\langle 0 | W^\dagger(0)
\delta\left(\eps- \int d \vec n\, (1-|\cos\theta|)\; {\cal E}(\vec 
n) \right)
W(0) |0\rangle_{k_\perp<\mu_{\rm IR}}\, ,
\label{f_1minusT}
\eea
with $\theta$ the angle between $\vec n$ and the thrust axis.
We define the operators $W$ in terms of path-ordered
nonabelian phase operators,
\bea
W(0)= P\;{\rm e}^{ig\int_0^\infty d\lambda \beta\cdot A(\lambda\beta)}\ 
\left[\, P\;{\rm e}^{ig\int_0^\infty d\lambda \beta'\cdot
A(\lambda\beta')}\, \right]^\dagger\, ,
\label{Wzero}
\eea
 and  the 
 operators ${\cal E}$ are defined to measure energy flow, by \cite{kos}
\bea
{\cal E}(\vec n)|N\rangle \equiv
\sum_{i=1}^N\delta(\cos\theta-\cos\theta_i)\; 
\delta(\varphi-\varphi_i)\; E_i|N\rangle\, ,
\label{operators}
\eea
for any final state with $N$ particles.
The matrix element in Eq.\ (\ref{f_1minusT})
 is matched to perturbation theory
by a cutoff in the transverse momentum of the soft
radiation at $\mu_{\rm IR}$.
 In this formulation, true universality resides 
at the level of correlators of energy flow in
the presence of the color sources,
\bea
{\cal G}(\vec n_1\dots \vec n_L;\mu)
= \langle 0 | W^\dagger(0)
{\cal E}(\vec n_1)\dots {\cal E}(\vec n_L)   W(0) |0\rangle\, .
\label{correlators}
\eea
A  ``mean field approximation'', which eliminates
nontrivial correlations between measurements
of energy flow in different directions:
\be
{\cal G}(\vec n_1\dots \vec n_L;\mu)
\to
\prod_{i=1}^L {\cal G}(\vec n_i,\mu)
\ee
reduces the shape function in (\ref{shape}) to
a delta function,
$f_e(\epsilon) \to \delta(\epsilon - \lambda_e)$.
In this approximation, the shift (\ref{shifteq})
is recovered.

We may take another viewpoint of
Eq.\  (\ref{f_1minusT}), 
and interpret it  as a matrix element in an
effective theory for soft radiation from
light-like color sources.
This is a natural language for the community with special 
interest in
inclusive B decay near the edge of phase
space, i.e., with jet-like final states. \cite{KSbFL}
In this case, the color source is the light
quark that emerges from the decay  $b\rightarrow s\gamma$,
for example, whose bremsstrahlung factorizes from
the remainder of the process.

\subsection{Exclusive B Decay}

Our third example is the recent application of
factorization to the fully exclusive B decays
into two mesons, $M_1,\, M_2$, 
either heavy-light ($D \pi$), or light-light
($\pi\pi,\, K\pi$).  Because the incoming quark is
heavy, the appropriate factorization 
is somewhat different than in the previous examples,
but the light-like relative velocity
of the light meson(s) in the final state leads
to a factorized form
for the decay amplitude ${\cal A}$ (here for the light-light case): 
\cite{BBN,Li}
\bea
{\cal A}(B\rightarrow M_1M_2)
&=&
F_{B\rightarrow M_1} \hat T^I(m_b,\mu)\; \otimes \Phi_{M_2}(\mu) 
\nonumber\\
&\ & \hspace{5mm} +
\hat T^{II}(m_b,\mu) \otimes \Phi_B(\mu)\;\otimes 
\Phi_{M_2}(\mu)\; \otimes \Phi_{M_1}(y,\mu)\, ,
\eea
where the short-distance functions $\hat T$ are computable in
perturbation theory, while the $\Phi$'s are nonperturbative
matrix elements that are wave functions for the hadrons.  
  Further
nonperturbative information is contained in $F_{B\rightarrow M_1}$,
which is itself a matrix element.
It may be possible to compute this matrix element
if transverse momenta are included in the convolution. \cite{Li}
Excitement has been generated by the possibility
of using the formalism to isolate weak,
CP-violating phases in these decays.  This should
be possible 
because all strong-interaction phases are contained in the functions $\hat T$,
to leading power in $m_b$.
We still have things to learn
about the relationship between the different approaches to this
factorization, and about the important role of power corrections.
Nevertheless, the extension of
factorization methods to this class of
physical problems, whose interest transcends QCD, is an important
step forward.

\section{BFKL and High Parton Density}

Everything we've discussed so far has involved 
hard scattering, and hence is restricted to rare
processes.  In recent years, however, considerable 
attention has returned
to the bulk of the high energy cross sections, involving
relatively low momentum transfers at high energy,
including the Regge limit ($s\rightarrow \infty$ with
$t$ fixed), diffractive scattering and the total cross
section.  This classic constellation of topics is coming
to the fore once again, in the light of the copious
HERA data on small-$x$ DIS, and 
renewed progress in the perturbative description of
the total cross section, the so-called ``perturbative
pomeron'' of QCD, as described by the celebrated BFKL equation.

\subsection{BFKL 2000}

The BFKL equation, with the LO kernel shown explicitly, can be
written as
\bea
\xi{d \psi(\xi,k_T)\over d\xi} &=& -{\alpha_s N\over \pi^2}\, \int 
{d^2k_T\over (k_T-k_T')^2}\;
\left[\; \psi(\xi,k_T) - {k_T'{}^2\over 2k_T^2} 
\psi(\xi,k_T')\; \right] +{\rm NLO}\, .
\label{bfkl}
\eea
The stimulus for much recent work is the newly-calculated explicit
form of the NLO kernel, the fruit of
a decade of effort. \cite{NLO}

As shown in \cite{bal96}, Eq.\ (\ref{bfkl})
may be derived from a factorization characteristic of
the Regge limit. \cite{sen83}  
To be specific, we consider
the forward scattering amplitude
for virtual-photon-proton scattering, which is related by the
optical theorem to the structure functions of DIS.
At low $x$, we need only include (color singlet) gluon exchanges
in the $t$-channel, or equivalently only the gluon distribution
$G(x,Q)$.  
Introducing the ``unintegrated'' gluon distribution 
$\psi(\xi,k_T)$ through
\bea
G(\xi,Q)&=&\int^Q{d^2k_T}\; \psi(\xi,k_T)\, ,
\label{psidef}
\eea
the relevant factorization for structure function $F$ is
\bea
F(x,Q^2) &=& \int d^2k_T\ c\left({x\over \xi},Q,k_T\right)\, 
\psi(\xi,k_T)\left(1+ {\cal O}\left ({1\over\ln^2x}\right )\right)\, ,
\label{altfact}
\eea
in terms of a modified coefficient function, $c$.
Relative to Eq.\ (\ref{fact}), the
roles of  $k_T$ and $\xi$ 
have been exchanged: $\xi$ is now the factorization scale,
separating, in the terminology of \cite{bal96} ``fast''
from ``slow'' quanta, and $k_T$ is now the convolution
variable.  On the one hand, the incoherence
of the dynamics of fast and slow quanta make it possible to
factorize the amplitude; on the other hand, Lorentz invariance
leaves the division between the two arbitrary.
This arbitrariness leads to an evolution equation,
the BFKL equation.  We note, however, that
the factorization in Eq.\ (\ref{altfact}) holds to 
next-to-leading logarithm in $x\sim s/Q^2$ only.
Beyond this level, we must generalize the equation itself.

The ansatz
$\psi \sim x^{-\omega}\; \left({k_T^2/\mu^2}\right)^{\gamma-1}$,
 in Eq.\ (\ref{bfkl}),
gives a consistency equation relating the exponents $\omega$
and $\gamma$.  With $\bar\alpha_s\equiv N_c\alpha_s/\pi$, this is
\be
\omega(\gamma)= \bar\alpha_s\chi_0(\gamma)\; \left[1-\beta_0{\alpha_s\over 4\pi}\, 
\ln{k^2\over\mu^2}\right]+\bar\alpha_s^2\chi_1(\gamma)\, ,
\ee
where the function $\chi_0(\gamma)$
has long been known, and where $\chi_1$ is new, and
the subject of much investigation.
 
 The largest value of $\omega$
gives the dominant small-$x$, or equivalently large-$s$
behavior. \cite{NLO} Exhibiting
only the LO result in analytic form, one finds
\be
\omega_{\rm max}=4N_c\ln2(\alpha_s/\pi)
[1-6.5\bar\alpha_s] \quad \Rightarrow \psi 
\sim 
s^{4N_c\ln2\;(\alpha_s/\pi)-{\rm large}}\, .
\ee
From LO in the kernel we have QCD Regge behavior, but the innocuous-looking
NLO result is, as it stands, not quite acceptable.  It is simply too 
large and negative, and can eventually lead, not only
to a decrease with $s$, but even to negative cross sections.
This produced a bit of initial consternation
on the part of some enthusiasts, but, ever-resourceful,
investigators have developed
very plausible proposals on
how to proceed.  
In fact, the problem may be traced to ``collinear divergences''
in $\chi_1(\gamma)$, which, from the limits $k_T\rightarrow 0$
and $k_T\rightarrow k^\prime_T$ in Eq.\ (\ref{bfkl}), receives
poles up to 
$\gamma^{-3}$ and $(1-\gamma)^{-3}$. \cite{salam}

 Proposals on how to interpret the NLO kernel have included:
(1)  Adjust the scale of $\alpha_s$, \cite{Betal}
(2) Impose kinematic constraints in $\cal K$, 
 demanding strong ordering of particles in rapidity; \cite{rapcut} 
(3)  Import information from DGLAP evolution,
given the association of collinear logarithms
in DIS to the singular behavior of $\chi$. \cite{CCABF} 
Particularly for the latter proposals, 
the connection of BFKL to small-$x$ DIS may suggest phenomenological
tests of their efficacy.  This story is probably just beginning.

\subsection{Effective theories and high parton density}

Sometimes it can be difficult for those not working on 
small-$x$ and BFKL to appreciate fully their perennial fascination.
One way of looking at 
what's special about BFKL evolution is 
that, if the LO BFKL equation is not too
misleading, then as we evolve to low $x$ we are
forced to a regime of 
high parton density even at
``fixed" (actually diffusing) virtuality.   
One dramatic manner
of thinking about this regime is as a strong-field
configuration of QCD, a
dense phase of weakly-interacting gluons.
\cite{cglass}  It may even be possible to bring
such a state into being in the laboratory; the RHIC
at Brookhaven may produce it
as an initial state in nuclear collisions, 
as may the LHC operating with nuclear beams.

This viewpoint has been developed quantitatively
through an effective theory, which has some similarities to the
one described above in the context of shape functions.
We introduce a set of color sources, this time
coming from the distant past along the lightcone,
\be
W_\pm(x^\mp,x_t) = P \exp\, \left[ \int_{-\infty}^\infty dx^\pm
 A^\mp(x^\mu)\right]\, .
\ee
The relation of 
the BFKL equation to such
an effective field theory was described in \cite{bal99}.
The nuclear connection is made by modelling
 a large nucleus as a  distribution of
the sources: \cite{Venu}
\bea
S_{\rm nuclear\ field}
=
S_{\rm QCD} + {i\over N_c} \int d^2x_tdx^-\rho(x_t,x^-) W_+(x^-,x_t)\, .
\eea
Among the intriguing results of this approach is the generation,
\cite{KrVe}
for a nucleus of essentially unlimited size, of
a  gluon occupation number density,  which is
of nonperturbative magnitude,  ${\cal O}(1/\alpha_s)$,
and which can serve as a starting point for the
very complex time evolution of nucleus-nucleus collisions.  
\cite{initial}

\section{Conclusions}

Even within the area of factorization at high energy, I have
of necessity passed over many developments from the past few years,
regarding global PDF fits \cite{pdf} and their uncertainties, \cite{uncer}
cross sections at measured transverse
momentum, \cite{qtpt} diffraction, \cite{diff}  higher-twist \cite{ht},
polarized \cite{pol} and skewed \cite{skew} 
parton distributions, and more.
The subject of power corrections at high
energy is still new, and we are just now learning
to read the quantum mechanical history of
QCD scattering in the language of final states. 
I expect the progress of the past few years, 
punctuated as it is with novel ideas and applications,
to continue for some time, as we ask new
questions of quantum chromodynamics.

\section*{Acknowledgments}
I would like to thank the organizers of the {\it IVth Rencontres
du Vietnam} for the chance to participate in a memorable 
conference, and to the National Science
Foundation for support. I would also like to thank
Brookhaven National Laboratory for its
hospitality.   This work was also supported in part 
by the  National Science Foundation,
grant PHY9722101.

%\section*{References}

\end{document}